\documentclass[prl,twocolumn]{revtex4-1}

\usepackage{amsmath}
\usepackage{bbm}
\usepackage{amssymb}
\usepackage{graphicx}
\usepackage{hyperref}
\usepackage{times}
\usepackage{color}
\newcommand{\je}[1]{\textcolor{black}{#1}}

\newcommand{\ii}{{\rm i}}
\newcommand{\e}{{\rm e}}
\newcommand{\ZZ}{\mathbb{Z}}
\newcommand{\RR}{\mathbb{R}}
\newcommand{\CC}{\mathbb{C}}
\newcommand{\TT}{\mathbb{T}}

\newcommand{\Imag}{{\rm Im}}
\newcommand{\Real}{{\rm Re}}

\def\one{\mathbbm{1}}
\def\b{{\rm b}}
\def\f{{\rm f}}
\def\r{{\rm r}}

\newcommand{\tr}{\operatorname{tr}}

\begin{document}

\title{Noise-driven quantum criticality}

\author{ J.\ Eisert$^{1}$ and T.\ Prosen$^{1,2}$}
\affiliation{$^1$ Institute of Physics and Astronomy, University of Potsdam, 14476 Potsdam, Germany}
\affiliation{$^2$Department of Physics, FMF, University of Ljubljana, 1000 Ljubljana, Slovenia}
%\affiliation{$^3$ Institute for Advanced Study Berlin, 14193 Berlin, Germany}

\begin{abstract}

We discuss a notion of quantum critical exponents in open 
quantum many-body systems driven by quantum noise. We show that 
in translationally invariant quantum lattice models undergoing quasi-local Markovian dissipative processes, 
mixed states emerge as stationary points that show scaling laws for the divergence of 
correlation lengths giving rise to well-defined critical exponents. 
The main new technical tool developed here is a complete description of 
steady states of free bosonic or fermionic translationally invariant systems driven by quantum noise:
This approach allows to express all correlation properties in terms of a symbol, paralleling
the Fisher-Hartwig theory used for ground state properties of free models. We discuss critical
exponents arising in bosonic and fermionic models. Finally, we 
relate the findings to recent work on dissipative preparation of pure dark and 
matrix product states by Markovian
noisy processes and sketch further perspectives.
\end{abstract}
\maketitle

Quantum phase transitions (QPT) and quantum criticality are among the most remarkable
emergent features and properties of large quantum quantum lattice systems. Even if
individual constituents directly interact only to their immediate
neighbors, characteristic length scales associated with 
two-body correlation functions diverge at critical points. Critical phenomena are 
responsible for a sizable number of 
effects in quantum many-body physics \cite{Sachdev,Vojta}. Specifically, if 
the Hamiltonian is of the form
\begin{eqnarray}\label{1}
	H_0= \sum_j \tau_j (h),
	\label{eq:ham}
\end{eqnarray}
where $h$ is a Hamiltonian term acting on a small number of sites only---usually nearest 
neighbors---and $\tau_j$ is the shift to the lattice site $j$, and $V$ is another Hamiltonian of this form, then
a system is critical at $g=g_c$
%undergoes a quantum phase transition at $g=g_c$
if the ground state energy of
$
	H=H_0+ g V
$
is non-analytical at $g_c$. In second order QPT this is
accompanied with a divergence of correlation lengths giving rise to the familiar notion of a critical point in quantum many-body systems. 

In this work, we quantitatively explore the possibility of scaling laws of correlation lengths of quantum
many-body systems driven by quantum noise.
%Center stage will be taken by 
%free models, for which we develop a theory of criticality paralleling the theory of criticality
%of ground states. 
The Hamiltonian will remain entirely unchanged, merely
we add Markovian noise that acts quasi-locally, so either on single sites alone or on a finite number
of neighbors. We will see that despite this local action of the dissipative dynamics, {\it mixed states} arise
as steady states that show characteristic divergences of correlation lengths, even allowing for the 
meaningful definition of critical exponents. The observation that local noise alone can be used to prepare 
pure quantum many-body states exhibiting long-range order was made in the seminal work of Refs.\ 
\cite{KrausLong,Zoller,Wolf,DiehlLong}. Here, we go beyond that insight by developing a theory of critical exponents
of (generically non-pure) non-equilibrum steady states of local dissipative systems. 
The main new technical tool will be a description of free 
translationally invariant bosonic and fermionic open quantum
systems driven by Markovian quantum noise---paralleling the central role free models play in the 
common theory of QPT. This approach allows to express correlation functions
of canonical coordinates in terms of a symbol reflecting a momentum space description,
reminding of the so powerful Fisher-Hartwig formalism  \cite{FH}
for ground state properties.

{\it Noise-driven critical exponents.} As far as the Hamiltonians are concerned, we study the---in the
theory of criticality---usual 
local quantum many-body systems on a lattice. The Hamiltonian can hence be written as in 
Eq.\ (\ref{1}); this is nothing but the familiar general translationally invariant 
local lattice model.
In addition to the local Hamiltonian, we allow for noisy processes merely {\it locally} affecting the 
quantum many body system. This noisy process is supposed to be Markovian. Hence,
each site $j$ is associated with a generator of  form
\begin{eqnarray*}
	{\cal L}_g (\rho) = \frac{1}{2}   \sum_\mu \left(
	 2 L_{\mu}(g)\rho L_{\mu}(g)^\dagger - 
	\{ L_{\mu}(g)^\dagger L_{\mu}(g) ,\rho \} 
	\right),
\end{eqnarray*}
where $g$ is a parameter in the many-body model. The Lindblad operators $\{L_{\mu}(g)\}$
in turn act merely on {\it finitely many sites}, typically on-site or nearest neighbors.
It is key to this work that noisy processes with long-range interactions are excluded---and still
we will find divergent correlation functions.
The equations of motion of the state are given by	
\begin{eqnarray}
	\frac{{\rm d}}{{\rm d}t}\rho = -\ii\sum_j [\tau_j (h),\rho] +\sum_j\tau_j( {\cal L}_g (\rho))
	\label{eq:eom}
\end{eqnarray}
where the summations ($j$) run over the $d$-dimensional lattice $\ZZ^d$.
For local observables $\{O_j\}$, where $O_j$ is supported on site $j$ only---again as usual--- one
considers correlation functions
\begin{eqnarray}
	f(j,k)=  |\langle O_j O_k\rangle - \langle O_j \rangle \langle O_k \rangle | 
	\label{eq:cor}
\end{eqnarray}
and defines the {\it correlation length} as the length scale
\begin{eqnarray}
	\xi= - \!\!\lim_{|j-k|\rightarrow\infty}\!\!\log f(j,k)/|j-k|.
\end{eqnarray}
In sharp contrast to Hamiltonian standard descriptions, we now 
put into the center of attention 
the dependence of steady states in terms of the noise only,
while keeping the Hamiltonian unchanged. As the main focus of this work, 
we will see that it makes sense to define a {\it critical value} $g_c$
of this parameter, such that when approaching 
\begin{eqnarray}
	g\rightarrow g_c
\end{eqnarray}	
the correlation length will diverge. In fact, one can define a 
{\it noise-driven critical exponent} $\lambda$, by means of 
the asymptotic behavior of this correlation length
\begin{eqnarray}
	\xi^{-1} \sim \Lambda|g-g_c|^\lambda
	\label{eq:crit}
\end{eqnarray}
for some $\Lambda>0$.
Note that while this parameter is defined as the critical exponent for 
critical phenomena in the usual sense, all divergences of correlation lengths
are entirely driven by quantum noise in this open quantum system.

{\it Free driven models.} Centre stage will be taken by free models, so models where the Hamiltonian
$h$ is quadratic in bosonic or fermionic operators and $L_\mu(g)$ are linear. 
Bosonic and fermionic systems will be treated in an as parallel as possible
fashion \cite{Statistics}.
Bosonic operators $\{b_j\}$ associated with site $j$
are accompanied by commutation relations, whereas fermionic operators $\{f_j\}$
anti-commute, 
\begin{eqnarray}
	[b_j,b_k^\dagger]= \delta_{j,k},\quad \{f_j,f_k^\dagger\}= \delta_{j,k}.
\end{eqnarray}
Suppose for a moment we ignore the translational invariance and consider a general 
quasi-free many body system defined on $n$ sites, where the limit $n\to\infty$ will be considered later.
The canonical variables can be grouped into a $2n$ dimensional vector 
of positions and momenta for bosons
$u=(u_{\nu,j};\nu=1,2;1\le j\le n)$, where
$u_{1,j} = b_j+b^\dagger_j$, $u_{2,j}= \ii (b_j - b_j^\dagger )$ 
and analogously of Majorana operators $w$ for fermions
%$w=(w_{\nu,j};\nu=1,2;1\le j \le n)$ 
with $w_{1,j} = f_j+f^\dagger_j  $ and 
$ w_{2,j}= \ii (f_j - f_j^\dagger)$. This convention, different from the usual one by a factor of $\sqrt{2}$, 
simplifies some of the later expressions.
%
% Check
%
%
% 

We (i) assume the Hamiltonian (\ref{eq:ham}) to be defined by 
a quadratic form, $H= u^{\rm T} H_\b u$, where $H_\b=H_\b^{\rm T}\in \RR^{2n\times 2n}$ 
for bosons, and $H = w^{\rm T} H_\f w$ where
$H_\f = -H_\f^{\rm T}\in \CC^{2n\times 2n}$, purely imaginary, 
matrix for fermions. (ii) The Lindblad operators $L_\mu = l_\mu^{\rm T} u$ ($L_\mu = l_\mu^{\rm T} w$) are assumed to be
linear in the coordinates and can be
written in terms of $2n$ dimensional vectors $l_\mu$, which can be grouped together into a Hermitian {\em bath matrix}
\begin{eqnarray}	
	M = \sum_\mu l_\mu \otimes \bar{l}_\mu,
	\label{eq:bath}
\end{eqnarray} 
$M\in \CC^{2n\times 2n}$, 
having a symmetric or anti-symmetric 
real or imaginary part, $M_{\rm r}=(M+\bar{M})/2=M_{\rm r}^{\rm T}$ and $M_{\rm i}=- \ii(M-\bar{M})/2
=-M_{\rm i}^{\rm T}$, respectively.
The elementary two-point correlations are gathered in a Hermitian covariance matrix, 
being a real, symmetric matrix for bosons $\gamma_\b\in \RR^{2n\times 2n}$,
%$\gamma_\b= \gamma_\b^{\rm T}$, with entries
%
%
% Check
%
\begin{eqnarray*}	
	(\gamma_\b)_{j,k}= \frac{1}{2}\tr \rho (u_j u_k + u_k u_j) = 
	\Real \tr (\rho   u_j  u_k), 
\end{eqnarray*}
and a real anti-symmetric matrix for fermions $\gamma_\f\in \RR^{2n\times 2n}$,
%
% Check
%
%
\begin{eqnarray*}
	(\gamma_\f)_{j,k} =  \frac{\ii}{2} \tr  \rho   (w_j  w_k - w_k  w_j) = -\Imag \tr (\rho w_j w_k ).
%	:= -\ii(\tr \rho (w \otimes w) - \one_{2n}) = -\gamma_\f^{\rm T}.
\end{eqnarray*}
Obviously, if one adds terms linear in bosonic  operators 
to the Hamiltonian \je{(fermionic first moments are always vanishing)}, 
then one should subtract the possibly non-vanishing mean values in the definition of the covariances.
%, which are otherwise vanishing. 
Defining a symplectic unit matrix $\sigma = \ii \sigma_{\rm y} \otimes \one_n$, the bosonic covariances satisfy the uncertainty relations in the form
$\gamma_\b + \ii \sigma \ge 0$, whereas an analogous relation for the fermionic covariances reads $\gamma^2_\f + \one_{2n} \ge 0$.

From the equations of motion of the canonical coordinates or the Majorana operators 
in the Heisenberg picture, one finds after a tedious but straightforward 
computation that 
the covariance matrix satisfies a closed set of equations of motion 
\begin{eqnarray}
\frac{{\rm d}}{{\rm d}t} \gamma_\eta = X_\eta^{\rm T} \gamma_\eta + \gamma_\eta X_\eta - Y_\eta, \quad \eta = \b,\f,
\label{eq:eom1}
\end{eqnarray}
taking the same form for bosons and fermions. The specific matrices governing this equation 
of motion are given by
\begin{eqnarray}
X_\b = \sigma (2 H_\b + 2 M_{\rm i}), \quad Y_\b = 4 \sigma^{\rm T} M_{\rm r} \sigma = Y_\b^{\rm T}
\label{eq:XYb}
\end{eqnarray}
for bosons and 
\begin{eqnarray}
X_\f = -2\ii H_\f + 2 M_{\rm r},\quad   Y_\f = 4 M_{\rm i} = -Y_\f^{\rm T},
\label{eq:XYf}
\end{eqnarray}
for fermions, where  $X_{\b},X_\f,Y_\b,Y_\f\in\RR^{2n\times 2n}$. 
Note an important distinction between bosonic and fermionic quasi-free semigroups:
Namely the spectrum of fermionic $X_\f$ (which determines all the decay rates of Liouvillean 
relaxation \cite{3Q,SpecLind}) can only lie on the positive real-side of the complex plane (since $M_\r \ge 0$, see Lemma 2.3 of Ref.\
\cite{SpecLind}), whereas in the bosonic case the spectrum of $X_\b$ (\ref{eq:XYb}) (which again determines all the Liouvillean decay rates \cite{3Qbos}) is not constrained, so the Liouvillean evolution becomes unstable---describing indefinite 
pumping-in of energy---when some of the eigenvalues of $X_\b$ attain negative real part.
\je{It is also easy to identify pure steady states, corresponding to the {\it dark states} of
Refs.\ \cite{Zoller,KrausLong}, but here applied to the free setting: They have covariance matrices 
for which all eigenvalues of $(\sigma\gamma_\b)^2$ or of $(\gamma_\f)^2$
are all equal to $1$ in the bosonic and fermionic case, respectively \cite{DS}.}

{\it Translationally invariant systems and symbols.}
Next, we consider explicit translational invariance on $j \in \ZZ^d_L$, so $n=L^d$,  
and subsequently take the limit $L\to\infty$. \je{In case of finite systems, the yet to be defined symbol
will be defined over a discrete {\em quasi-momentum} space and the analysis does not apply: Hence, in all that follows, divergences of correlation
lengths manifest genuine quantum many-body effects.}
Due to translational invariance we may write
\begin{eqnarray*}
	(H_\eta)_{(\nu,j),(\nu',j')} = \{h_\eta (j-j')\}_{\nu,\nu'},\,\,\nu,\nu'=1,2,
\end{eqnarray*}
and similarly for the Lindblad operators $L_\mu$ introducing vectors $(l_\mu)_\nu$ on $\ZZ^d_L$.
Then we may define {\em the Hamiltonian and the Lindbladian symbols}, as, respectively, $2\times 2$ matrix valued and $2$-dimensional vector valued
functions on $\TT^d = [-\pi,\pi)^d \ni \varphi$
\begin{eqnarray*}
 	\tilde{h}_\eta (\varphi)  &=& \sum_{j\in\ZZ^d}  h_\eta(j)  \e^{-\ii\varphi\cdot j}, \,
	(\tilde{l}_{\mu})_{\nu} (\varphi) = \sum_{j\in\ZZ^d} (l_{\mu})_{(\nu,j)} \e^{-\ii\varphi\cdot j}.
\end{eqnarray*}
Note that $l_\mu$ here denotes only those Lindblad vectors which directly couple to the site $j=0$ of the lattice, all the other Lindblad vectors are obtained by the translations (as in Eq.\ (\ref{eq:eom})). By the convolution theorem, the bath matrix $M$ of Eq.\ 
(\ref{eq:bath}) is then also circulant, with a symbol
\begin{eqnarray*}
	\tilde{m}(\varphi) = \sum_\mu \tilde{l}_\mu(\varphi) \otimes \overline{\tilde{l}_\mu(\varphi)},
\end{eqnarray*}
which will be, for convenience, decoupled into a real and imaginary part
$\tilde{m}_{\rm r}(\varphi) = (\tilde{m}(\varphi)+\tilde{m}^{\rm T}(-\varphi))/2$ 
and $\tilde{m}_{\rm i}(\varphi) = -\ii(\tilde{m}(\varphi)-\tilde{m}^{\rm T}(-\varphi))/2$, respectively.

Furthermore, if the stationary point of the dynamical semigroup (\ref{eq:eom}) is unique, it is also 
translationally invariant: This follows immediately from a group twirl with shifts $\tau_j$ forming a
unitary representation: The steady state covariance matrix $\gamma_\eta$ is also (block) circulant. 
Then the condition for the fixed point of (\ref{eq:eom1}), namely the Lyapunov-Sylvester equation \cite{Sylv} 
$X_\eta^{\rm T} \gamma_\eta + \gamma_\eta X = Y_\eta$ 
maps to a simple $2\times 2$ matrix equation for the {\em covariance symbol}
\begin{eqnarray}
	\{\tilde{\gamma}_\eta(\varphi)\}_{\nu,\nu'} = \sum_{j\in\ZZ^d} \{\gamma_\eta\}_{(\nu,j),(\nu',0)} 
	\exp(-\ii\varphi\cdot j),
	\label{eq:tisylv}
\end{eqnarray}
namely
\begin{eqnarray}
\tilde{x}^{\rm T}_\eta (-\varphi) \tilde{\gamma}_\eta(\varphi) + \tilde{\gamma}_\eta(\varphi)\tilde{x}_\eta(\varphi) = \tilde{y}_\eta(\varphi)
\end{eqnarray}
where for bosons $\tilde{x}_\b = \sigma_2 (2\tilde{h}_\b + 2\tilde{m}_{\rm i}), \tilde{y}_\b = 4 \sigma_2^{\rm T} \tilde{m}_{\rm r}\sigma_2$ (where $\sigma_2=\ii\sigma_{\rm y}$), and
for fermions $\tilde{x}_\f = -2\ii \tilde{h}_\f + 2\tilde{m}_{\rm r},  {\tilde y}_\f = 4 \tilde{m}_{\rm i}$ in direct consequence of (\ref{eq:XYb},\ref{eq:XYf}).
We note that Eq.\ (\ref{eq:tisylv}) is a simple $2\times 2$ Sylvester matrix equation (in fact a linear system of $4\times 4$ equations for matrix elements of $\tilde{\gamma}_\eta$)
for a given fixed quasi-momentum $\varphi$, and can be trivially solved.
The real-space two-point correlation function (\ref{eq:cor}) is then obtain by inverse Fourier  transforming the covariance symbol
\begin{eqnarray}
	\gamma_\eta (r) = \frac{1}{(2\pi)^d} \int_{\TT^d} 
	{\rm d}^d \varphi\, \tilde{\gamma}_\eta(\varphi)\exp(\ii \varphi\cdot r).
	\label{eq:mainequation}
\end{eqnarray}

{\it One-dimensional models.}
Clearly, for systems with short-range or finite-range interaction, the symbols $\tilde{h}_\eta$ and $\tilde{m}$, and hence $\tilde{x}_\eta$,
are {\em holomorphic} on $\CC^d $, therefore the solution $\tilde{\gamma}_\eta(\varphi)$ of (\ref{eq:mainequation}) can have at most simple pole singularities, 
$\varphi^* \in \CC^d$, determined by either of the following conditions
\begin{eqnarray*}
	\beta_\nu(\varphi^*) + \beta_\nu(-\varphi^*) = 0,
	\quad{\rm or}\quad \beta_1(\varphi^*) + \beta_2(-\varphi^*) =0
\end{eqnarray*}
 where $\beta_{1,2}(\varphi)$ denote the two eigenvalues of $\tilde{x}_\eta(\varphi)$.
For the rest of the discussion we focus on one dimension $d=1$.
Then, if $\varphi^*$ is an isolated pole singularity with minimal $|\Imag\,\varphi^*|$, 
then it takes straightforward complex analysis to show that $\exists C > 0$, such that 
%$\| \gamma_\eta(r) \| < C \exp(-|\Imag\varphi^*| |r|)$, hence the correlation length can be estimated as
%\begin{eqnarray*}
%	\xi \sim |\Imag\varphi^*|^{-1}.
%\end{eqnarray*}
%{\je{What is shown is something stronger: The above would only imply an upper bound. 
%What is rather true, or not, is that
%
%
%
% Check
%
%
\begin{eqnarray}
	\lim_{r\rightarrow \infty}\frac{\|\gamma_\eta(r)\| }{\exp(-|\Imag\varphi^*| |r|)} = C, \quad {i.e.}\;\; \xi^{-1} = |\Imag\varphi^*|.
\end{eqnarray}
This comprises the central result of our Letter. The criticality of the system is then 
signaled by the closest pole $\varphi^*$ approaching the {\em real} circle $\TT^1$.
 Clearly, due to continuity of $\tilde{x}(\varphi)$, this means that for a 
 Bloch quasi-momentum $\varphi'  = {\rm Re}\,\varphi^*$, the eigenvalue
of $\tilde{x}(\varphi)$, whose real part $1/\tau$ determines the relaxation time of the 
corresponding Liouvillean normal mode,
can be estimated as 
$1/\tau < C' |\Imag\,\varphi^*|$ where constant $C'$ may depend on system parameters but not on $\varphi^*$. 
Thus the criticality (\ref{eq:crit}) also implies a {\em critical slowing down} $\tau > C' \xi$.
Note that this way of finding critical points is different from the usual
approach in Hamiltonian problems, constituting a difference in how dissipative and non-dissipative systems
are treated. \je{Dynamical instabilities may occur and may be detected in this framework, but the
divergent correlation lenghts always reflects a true quantum many-body effect and can not be viewed as
local dynamical instabilities.}

For the fermionic case, it can also be shown that in case the pole arrives right down to the real torus $\TT^d$, i.e. $\Imag \varphi^* = 0$, then the solution of 
Sylvester equation there has to vanish $\tilde{\gamma}_\f(\varphi'=\varphi^*)=0$ 
(a consequence of Lemma 2.5 of Ref.\cite{SpecLind}), and hence algebraic decay of correlations is not achievable.    

{\it Example 1: Driven criticality for fermions.} We conclude the discussion of the quasi-free case by 
providing explicit examples. An interesting example in the fermionic case in one dimension ($d=1$) is provided by the fermionic model 
corresponding to a XY spin-$1/2$ chain with exchange couplings $J_{\rm x}=\frac{1}{2}(1+\Gamma),J_{\rm y}=\frac{1}{2}(1-\Gamma)$ and transverse magnetic field $B$, by virtue of the Jordan-Wigner 
transformation (see, e.g., Ref.\ \cite{Prosen} for the exact description of the model). 
The Hamiltonian symbol is $h_\f(\varphi) = \frac{1}{2}(B - \cos\varphi) \sigma_{\rm y} + \frac{\Gamma}{2} (\sin \varphi) \sigma_{\rm x}$.
We first discuss the case of a {\em spatially incoherent}, on-site noise, where a single Lindblad operator per site is considered
$L_j(g) = \varepsilon (f_j + f_j^\dagger) + \varepsilon \ii (f_j - f^\dagger_j)e^{\ii g}$, where we find 
\begin{eqnarray*}
	\xi^{-1}=\Imag\,\varphi^* \sim \varepsilon^2((|B|-1)^2 + \varepsilon^4\sin^2 g).
\end{eqnarray*}
This means that we may have criticality as $g\to g_c = 0$ or $g_c=\pi$, only if the (noise-free)
Hamiltonian XY model is already critical, namely if $|B|=1$.
Therefore, on-site noise cannot induce criticality. However, things change drastically 
when we consider 
two-site spatially coherent noise. For simplicity we again take a single Lindblad operator per site
$
	L_j(g) = \varepsilon (f_j + f_j^\dagger) + \varepsilon (f_{j+1} + f_{j+1}^\dagger) e^{\ii g}.
$
This term does not violate the parity of fermion number superselection rule \cite{F} and defines
valid and physically meaningful noise. 
In this case, our analysis yields a very simple expression for the covariance symbol
\begin{eqnarray*}
	{\gamma}_\f(\varphi) = \frac{ \sin g \sin \varphi}{2(1+ \cos g \cos\varphi )} \one_2
\end{eqnarray*}
which, quite astonishingly, does not depend on any of the hamiltonian parameters ($\Gamma,B$) at all!
Now, we find a clear, noise induced criticality for $g_c = 0$ or $g_c=\pi$, with the inverse correlation length $\xi^{-1} = {\rm arcosh}(1/\cos g)$, which close to the critical
point can be estimated as $\xi \sim |g-g_c|^{-1/2}$ 
giving the critical exponent $\lambda=1/2$, same as in the boundary-driven XY model \cite{Prosen}, and also
as for {\it ground states} of free bosons.
%Finding the mean field critical exponents (as in Ref.\cite{Prosen}) should not be surprising as we are dealing with non-interacting models.

{\it Example 2: Driven criticality for bosons.} 
As a second example we consider, for $t,v\in \RR$,
a simple nearest neighbor hopping translationally invariant bosonic chain, 
\begin{eqnarray*}
	H =t \sum_j (b_j^\dagger b_{j+1} + b_{j+1}^\dagger b_j - (v+2) b^\dagger_j b_j),
\end{eqnarray*}
%$t<0$,
with Hamiltonian symbol $h_\b(\varphi) = t (\cos\varphi - v)$,
and on-site noise with a single Lindblad operator per site
$
	L_j(g) =  \varepsilon (b_j + b_j^\dagger) + \varepsilon \ii (b_j - b^\dagger_j)e^{\ii g} 
$
in analogy to the previous fermionic case.
The eigenvalues of $X_\b$, $\beta_{1,2} = 2\varepsilon^2 \sin g \pm 4\ii t |v-\cos\varphi|^{1/2}$ signal {\em instability} of the Liouvillean fixed point 
(steady state) for $g \in (-\pi,0)$ where ${\rm Re}\, \beta_{1,2} < 0$.
Following previous paragraphs, the covariance symbol reads
\begin{eqnarray*}
	\tilde{\gamma}_\b(\varphi)=\begin{pmatrix}
	\frac{1}{\sin g} \!+\!\frac{2t \varepsilon^2 
	(v-\cos\varphi)\cos g}{z(\varphi)}\!\!\!\! & \frac{\varepsilon^4 \sin g\cos g}{z(\varphi)} \cr
	\frac{\varepsilon^4 \sin g\cos g}{z(\varphi)} & 
	\!\!\!\!\!\frac{1}{\sin g}\!-\!\frac{2t \varepsilon^2 (v-\cos\varphi)\cos g}{z(\varphi)}
\end{pmatrix},
\end{eqnarray*}
$z(\varphi)=4t^2(v-\cos\varphi)^2 + \varepsilon^4 \sin^2 g$. The pole singularity is determined by a zero of 
 $z(\varphi^*)=0$, $\varphi^* = \arccos(v \pm \ii \varepsilon\sin g/2t)$.  Two regimes emerge: (i) If one has an {\em optical gap} 
 $|v| > 1$, $\Imag\,\varphi^*$ is bounded from below independently of $g$ and no noise-induced 
 criticality occurs. (ii) In the {\em acoustic regime} $|v| \le 1$, yet, 
 we may expand $\Imag\varphi^*$ in $\sin g$ and obtain 
 \begin{eqnarray*}
 	\xi^{-1}=|\Imag\,\varphi^*| \approx \bigl(\varepsilon/(2t ({1-v^2})^{1/2}  )\bigr) |\sin g|,
\end{eqnarray*}
giving the critical points $g_c = 0,\pi$ and the critical exponent $\lambda=1$. We note, 
however, that noise induced criticality of a bosonic system corresponds exactly 
with the instability thresholds discussed above. A similar analysis can be implemented also for two-site coherent noise.
 
%It is also an easy exercise to see what is going on for spatially coherent two-site noise, 
%\begin{eqnarray*}
%	L_j(g) =  \varepsilon (b_j + b_j^\dagger) + \varepsilon (b_{j+1} + b^\dagger_{j+1})e^{\ii g}. 
%\end{eqnarray*}
%There we find, in the optical regime $|u| > 1$ that again noise induced criticality is always absent, whereas in the acoustic regime $|u| < 1$ we always find, for any noise parameters $\varepsilon,g$, that the system has both, always unstable steady state, and singularity $\varphi^* = {\rm arcccos} (u)$ is always real.
%  \tp{We should discuss this, to see whether to include anything about two-site bosonic noise...}
 
{\it Remarks on the relationship with matrix-product states as dark states.} The works
\cite{Zoller,KrausLong} consider pure states as dark states of dissipative processes. Indeed,
following the ideas of Ref.\ \cite{KrausLong}, quasi-local dissipative Markovian 
processes can be constructed with {\it matrix product states} \cite{MPSReview}
being the unique dark states. This insight can be combined with one of Ref.\ \cite{WolfPhaseTransition},
where parent Hamiltonians are constructed having unique ground states, and which undergo a quantum
phase transition of any order when parameters of the Hamiltonian are being altered. 
E.g., Ref.\ \cite{WolfPhaseTransition} considers 
$H=\sum_j \tau_j (h)$, where, with $S$ being the vector of spin-$1$ matrices, 
\begin{eqnarray*}
	h&=&  
	(2+g^2) (S \otimes S)
	+ 2 (S \otimes S)^2
	+ 2 (4-g^2) (S_{\rm z}\otimes 1)^2 \nonumber\\
	&+& (g+2)^2 (S_{\rm z} \otimes S_{\rm z})^2
	+ g(g+2) \{ S_{\rm z} \otimes S_{\rm z}, S\otimes  S\}
	,
\end{eqnarray*}
with unique ground state for $g\neq g_c=0$.
Following the construction of Ref.\ \cite{KrausLong}, one can then identify quasi-local dissipative 
processes -- in just the same sense as above --
giving rise to critical exponent $\lambda=1$ \cite{AS}.

{\it Steady states and entanglement area laws.} We finally discuss the entanglement 
structure of steady states driven by dissipation and identify a connection to
{\it entanglement area laws}.
This connection is particularly manifest in the case of free 
bosonic systems. Since their steady states are usually mixed, a genuine
entanglement measure, and not the entropy of a reduced state, 
has to be applied. We bound these entanglement measures of a region $A$
distinguished from the rest of a one-dimensional chain $L$, such as the {\it distillable entanglement}, 
from
above by the logarithmic negativity \cite{LN}, 
defined for a state $\rho$ as $E_N(\rho)= \log_2\|\rho^\Gamma\|_1$, where
$\rho^\Gamma$ denotes the partial transpose of $\rho$ with respect to the degrees of freedom of $A$. 
The state being Gaussian,
the covariance matrix of $\rho^\Gamma$
is given by
$\gamma_\b^\Gamma = P\gamma_\b P$, where $P= \one_n\oplus Q$ with $Q$ being the diagonal matrix 
whose main diagonal elements are $-1$ when a degree of freedom is associated with $A$ and $1$ otherwise.
The logarithmic negativity can in turn be computed from the
symplectic eigenvalues $\{\lambda_j\}$ of $\gamma_\b^\Gamma$, given by
the singly counted positive square roots of the eigenvalues of the matrix 
$K=(\sigma)^{1/2} \gamma_\b^\Gamma \ii \sigma \gamma_\b^\Gamma(-\sigma)^{1/2}$.
Using a bound reminding of one established in Ref.\ \cite{Area}, one finds
\begin{eqnarray*}
	\log_2 \|\rho^\Gamma\|_1 &=& \sum_j \log_2 \max(1, \lambda_j^{-1}) \leq 
	 \sum_j |\lambda_j^{-1}-1|\nonumber\\
	 &=&\| K^{-1/2}-\one_{2n}\|_1\leq \| K^{-1/2}-\one_{2n}\|_{l_1},
\end{eqnarray*}
where the norm $\|.\|_{l_1}$ 
adds all absolute values of all matrix entries.
From this, one can see that whenever entries of $\gamma_\b(r)$ are exponentially decaying away from the main diagonal---as they always do in the models considered here---one 
encounters an upper bound that does not depend on the number of sites of $A$. Hence all steady states of the considered one-dimensional bosonic models satisfy an entanglement area law.

{\it Summary and outlook.} In this work, we introduced the concept of critical exponents of 
quantum many-body systems driven by quantum noise. We identified scaling laws reminding of
the theory of criticality for ground states of local quantum many-body models. At the heart of
our discussion is a theory of driven free models---providing an important simple class of models
for which all features of criticality can be established analytically, reminding of the theory of ground
state criticality for free models. It is the hope that this work contributes to paving 
the ground for a systematic
study of preparation of complex states by dissipation and quantum noise.

{\it Acknowledgements.} We warmly thank S.\ Diehl for inspiring discussions and feedback on the 
manuscript and 
the EU (Qessence, Compas, Minos), the EURYI scheme, 
and the Alexander von Humboldt Foundation for support.


\begin{thebibliography}{99}
\bibitem{AS}
	This argument as such does not provide a bound to the spectral gap
	dictating the rate of convergence to this stationary point.
\end{thebibliography}
\end{document}